\documentclass[aps,prl,amsmath,amssymb,twocolumn,groupedaddress,
showpacs,nofootinbib]{revtex4-1}

\usepackage{graphicx} 
\usepackage{dcolumn} 
\usepackage{bm}
\usepackage{color}
\usepackage{ulem}

\begin{document}

\title{Direct observation of the injection dynamics of a laser wakefield accelerator using few-femtosecond shadowgraphy}

\author{A.~S\"{a}vert$^{1}$}
\author{S.~P.~D.~Mangles$^{2}$}
\author{M.~Schnell$^{1}$}
\author{E.~Siminos$^{3}$}
\author{J.~M.~Cole$^{2}$}
\author{M.~Leier$^{1}$}
\author{M.~Reuter$^{1,4}$}
\author{M.~B.~Schwab$^{1}$}
\author{M.~M\"{o}ller$^{1}$}
\author{K.~Poder$^{2}$}
\author{O.~J\"{a}ckel$^{4}$}
\author{G.~G.~Paulus$^{1,4}$}
\author{C.~Spielmann$^{1,4}$}
\author{S.~Skupin$^{5}$}
\author{Z.~Najmudin$^{2}$}
\author{M.~C.~Kaluza$^{1,4}$}

\affiliation{$^1$Institut f\"{u}r Optik und
  Quantenelektronik, Abbe-Center of Photonics, Friedrich-Schiller-Universit\"{a}t,
  07743 Jena, Germany} 

\affiliation{$^2$The John Adams Institute for Accelerator Science,
    The Blackett Laboratory, Imperial College London, London
    SW7 2AZ, United Kingdom.}

\affiliation{$^{3}$Max Planck Institute for the Physics of Complex Systems,
	01187 Dresden, Germany}
	
\affiliation{$^4$Helmholtz-Institut Jena,
  Friedrich-Schiller-Universit\"{a}t, 07743 Jena, Germany}

\affiliation{$^5$Univ. Bordeaux - CNRS- CEA, Centre Lasers Intense et Applications, UMR 5107, 33405 Talence, France}

\begin{abstract}
We present few-femtosecond shadowgraphic snapshots taken during the non-linear evolution of the plasma wave in a laser wakefield accelerator with transverse synchronized few-cycle probe pulses. 
These snapshots can be directly associated with the electron density distribution within the plasma wave and give quantitative information about its size and shape. Our results show that self-injection of electrons into the first plasma wave period is induced by a lengthening of the first plasma period. Three dimensional particle in cell simulations support our observations. 
\end{abstract}

\maketitle 

Laser-wakefield accelerators (LWFA) operating in the `bubble'-regime\,\cite{pukhov02} can generate quasimono\-ener\-getic multigiga\-electron\-volt electron beams\,\cite{leemans06,wang13} with femtosecond duration\,\cite{lundh11,buck11} and micrometer dimensions\,\cite{schnell12,plateau12}. These beams are produced by accelerating electrons in laser-driven plasma waves over centimeter distances. They have the potential to be compact alternatives to conventional accelerators\,\cite{hooker13}. In a LWFA, the short driving laser pulse displaces plasma electrons from the stationary background ions. The generated space charge fields cause the electrons to oscillate and form a plasma wave in the laser's wake. This  wave follows the laser at almost $c$, the speed of light; for low amplitude it has a wavelength of
  \begin{equation}
  \lambda_{\rm p} = 2\pi c\sqrt{\varepsilon_0 m_{\rm e}/(n_{\rm e}e^2)},
  \label{eq:one}
  \end{equation}
  where $n_{\rm e}$ is the electron  density of the plasma. At high amplitude, electrons from the background can be injected into the wake and accelerated, producing monoenergetic electron pulses\,\cite{mangles04,geddes04, faure04}. Significant progress has been made regarding achievable peak energy\,\cite{wang13}, beam stability\,\cite{gonsalves11} and the generation of bright X-ray pulses\,\cite{rousse04,kneip10,powers14}. Until now, most of our knowledge about the dynamics of the self-injection process has been derived from detailed particle-in-cell (PIC) simulations. These simulations show that self-focusing\,\cite{thomasPRL07} and pulse compression\,\cite{faurePRL05} play a vital role in increasing the laser pulse intensity prior to injection. Furthermore, simulations indicate that self-injection of electrons is associated with a dynamic lengthening of the first plasma wave's period (the `bubble'). This lengthening can be driven by changes of the electric field structure inside the plasma wave caused by the injected electrons\,\cite{tzoufras09}. In contrast, the lengthening may also be due to an intensity amplification of the laser pulse caused by the non-linear evolution of the plasma wave\,\cite{kalmykov09,kostyukov09} or due to a local increase in intensity caused by two colliding pulses\,\cite{lehe13}. In these latter scenarios, injection is a consequence of the lengthening of the bubble. However, experimental insight into these processes is extremely challenging due to the small spatial and temporal scales of a LWFA.

\begin{figure}
\includegraphics[width=246pt]{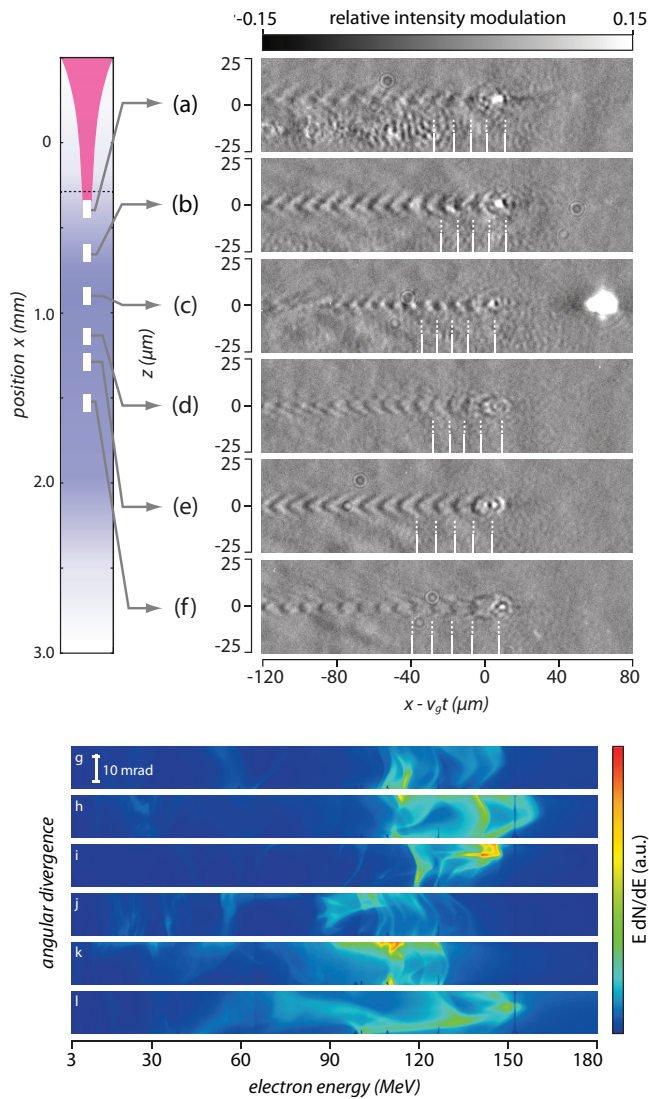}
\caption{\label{fig1}Left: gas/electron density profile and focus position (dashed line). a-f: Experimental shadowgrams at various positions in the plasma at a background electron density of $n_{\rm e}=1.65\times 10^{19}$cm$^{-3}$. The vertical lines indicate the on-axis position of the plasma wave's peaks as deduced from simulated shadowgrams (cf. Fig.\,\ref{fig3}). g-l: Energy in the electron beam per MeV and spatially resolved in the vertical coordinate corresponding to the  above images.}
\end{figure}

The plasma wave, a variation in the electron density, has an associated refractive index profile which can be detected using longitudinal\,\cite{marques96,siders96,matlis06} or transverse probes\,\cite{buck11}. Longitudinal probes cannot measure the rapid and dynamic evolution of the plasma wave that occurs in non-linear wakefield accelerators and suffer from the strong refraction caused by the steep refractive index gradients in a plasma wave. However, a small offset angle between pump and probe can measure the evolution, but only in 1D\,\cite{Li14}. Furthermore, to properly resolve the wake structure the probe must have a duration (or equivalent bandwidth) less then $\lambda_{\rm{p}}/c$ which has not yet been achieved\,\cite{buck11,Li14}, meaning that the important details of the wake evolution, e.g. the lengthening of the bubble in relation to the injection process have not yet been resolved. By using a probe pulse shorter than $\lambda_{\rm{p}}/c$ to perform high resolution shadowgraphy we are able to show for the first time that, under our experimental conditions, bubble expansion occurs before self-injection starts.

In the present study, the JETI-laser system at the Institut f\"{u}r Optik und Quantenelektronik in Jena, Germany delivered pulses of 750\,mJ energy and 35\,fs duration (central wavelength $\lambda_{\rm L}=810\,\rm{nm}$). The pulses were focused by an $f/13$ off-axis parabolic mirror to an elliptical focal spot with dimensions (FWHM) $8.9\,\mathrm{\mu m} \times 12.8\,\mathrm{\mu m}$ containing $27\%$ of the energy, resulting in peak intensities of $I_{\rm L}=6\times10^{18}\,$Wcm$^{-2}$ (corresponding to a peak normalized vector potential of $a_0=8.55\times 10^{-10}\lambda_{\rm{L}}(\mathrm{\mu m})\cdot\sqrt{I_{\rm L}(\rm{Wcm}^{-2})} \approx 1.7$). A supersonic helium gas jet was used, generating a plasma with electron density $n_{\rm e}$ in the range of $(0.5\ldots2.5)\times 10^{19}\,$cm$^{-3}$. Electrons accelerated during the interaction could be detected using a magnetic spectrometer or a scintillating screen. A small fraction of the laser was split from the main pulse, spectrally broadened in a hollow-core fiber filled with argon to support a transform limited pulse duration of $\tau_{\rm FL}=4.4\,$fs. Using dispersive mirrors and glass wedges to optimize dispersion, probe pulses as short as $\tau_{\rm probe}=(5.9\pm0.4)\,$fs were created\,\cite{schwab13}. These synchronized, few-cycle probe pulses were used to back-light the LWFA perpendicularly to the pump-pulse direction. A high-resolution imaging system produced shadowgraphic images with micrometer resolution on a CCD camera. By varying the delay between pump and probe, different stages of the plasma wave's evolution were recorded on subsequent shots close to the threshold density for self injection. The snapshots shown in Fig.\,\ref{fig1} are representative of each stage in the acceleration process. Shots were selected that exhibit similar quasi-monoenergetic electron spectra (Fig.\,\ref{fig1}(g-l)) and produced a high contrast shadowgram. The latter was affected by jitter in probe duration and pointing fluctuations of the pump which shifts the image out of focus. To reduce modulations induced by the probe pulse's beam profile, the relative intensity modulation was plotted using $I_{\rm norm}=(I-I_0)/I_0,$ with $I$ being the pixel value at each individual position and $I_0$ the value derived from a low order spline fit in the horizontal direction.

\begin{figure*}[!ht]
\includegraphics[width=\textwidth,clip=true]{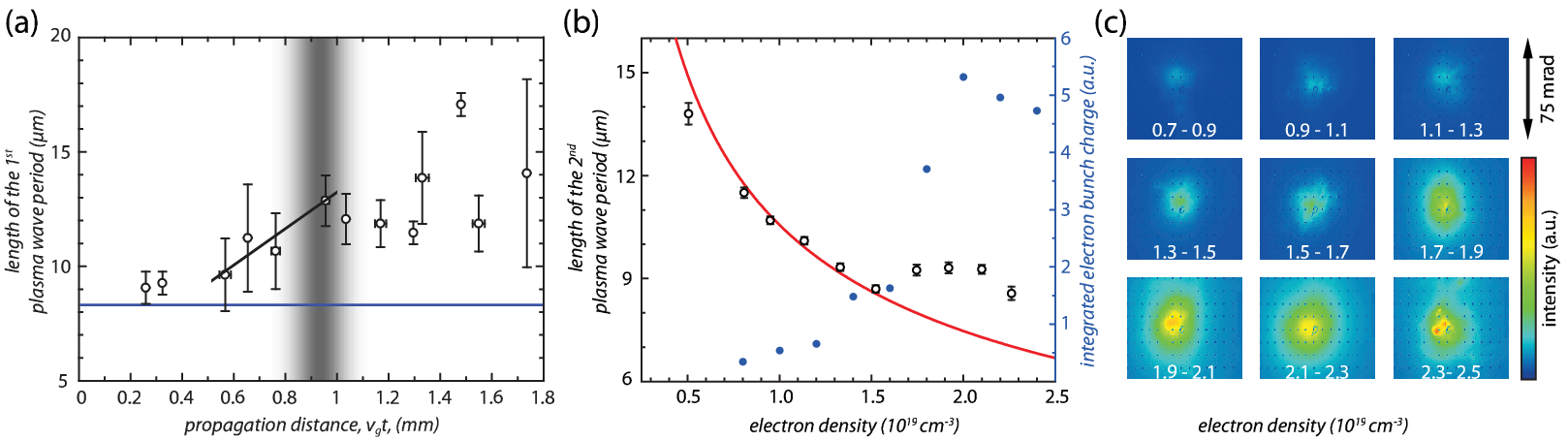}
\caption{\label{fig2}(a) Length of the $1^{\rm{st}}$ plasma period (bubble) as a function of propagation distance $v_{\rm g}t$ taken from the shadowgrams [error bars represent combination of uncertainty in determining the bubble length ($\pm0.5\,\mathrm{\mu m}$) in a single shot reduced by $\sqrt{N}$ where repeat shots are available]. Blue (solid) horizontal line is the expected $\lambda_p$ for $n_{\rm e} = 1.65\times10^{19}\,$cm$^{-3}$. Between $v_{\rm g}t =0.86$ and $1.0\,$mm (grey shaded area) wavebreaking radiation was frequently detected. (b) Wavelength of $2^{\rm{nd}}$ plasma period versus $n_{\rm e}$ at a fixed position $v_{\rm g}t =1.0\,$mm and corresponding integrated electron bunch charge. Open circles represent measured and averaged data points with the standard error of the mean and red line is according to Eq.~(\ref{eq:one}). (c) Electron beam profiles for different plasma densities averaged over 40-180 shots.}
\end{figure*}

Early in the interaction the regions of high and low electron density gradient (dark and light regions in the image) are approximately equal in length, indicating a linear plasma wave, Fig.\,\ref{fig1}(a). The transverse extent of the plasma wave reduces and the amplitude of the wave increases, Fig.\,\ref{fig1}(b). Later on, significant increase of curvature of the plasma-wave train and, in particular, in the lengthening of the first plasma period, Fig.\,\ref{fig1}(c) is apparent. Just ahead of the region where this lengthening starts we observe bright emission from the plasma. This has a broad spectrum (covering at least a range from 600- 1000 nm, cf. the 60 nm bandwidth of the pump) and is consistent with `wavebreaking radiation'\,\cite{thomas07}, which is a direct signature of the onset of self-injection in the experiment. Further propagation enhances the density gradient at the front of the bubble which now appears in the shadowgrams at the beginning of the wave train. After wavebreaking, the wave becomes highly non-linear, as indicated by the reversal in the direction of curvature of the trailing wave periods in the shadowgrams, Fig.\,\ref{fig1}(d) to \ref{fig1}(f). These features are closely linked to the process of transverse wavebreaking\,\cite{bulanov97}.

Our snapshots reveal that the dynamic process of bubble lengthening is intimately tied to self-injection. We plot in Fig.\,\ref{fig2}(a) the evolution of the plasma wave's first period. Early in the interaction the length of the first period has already increased as compared to the wavelength for a linear relativistic plasma wave, $\lambda_{\rm p}=2\pi c/\omega_{\rm p}$. The length of the bubble increases up to the point of wavebreaking, cf. Fig\,\ref{fig1}(c). During a single interaction, this radiation is emitted from a distinct spot on the optical axis with longitudinal position varying slightly in the experiment around $(930\pm67)\,\rm{\mu}$m for $n_e\approx1.65\times 10^{19} \mathrm{cm}^{-3}$. Beyond this point the shape of the plasma wave varies from shot to shot. This can  lead to the formation of a single bubble or to the merging of the first two plasma wave periods due to beam loading resulting in a strong variation of the bubble length after injection (Fig.\,\ref{fig2}(a)). 
A linear regression fit to the data (Fig.\,\ref{fig2}(a)) shows that the bubble starts expanding from a diameter of $(9.4\pm1.0)\,\mu \mathrm{m}$ at an expansion velocity of $v_{be}=(2.4\pm 1.4)\times 10^6\,\mathrm{ms}^{-1}$. The fact that we observe bubble lengthening before injection clearly demonstrates that the initial expansion is not caused by the effect of charge loaded into the wake. This lengthening is therefore most likely caused by intensity amplification of the pulse as it propagates in the plasma wave.

We also measured the length of the second wave period, shown in Fig.\,\ref{fig2}(b), as a function of density at a fixed position in the plasma ($v_{\rm g}t =1.0\,$mm). These measurements were made sufficiently far into the gas jet to ensure that it was in the uniform density plateau. At low densities, the length is well matched to Eq.~(\ref{eq:one}), but at high densities, $\lambda_{\rm p}$ is significantly longer. The density at which this transition occurs corresponds to the onset of injected charge, see Fig.\,\ref{fig2}(c), and to the self-injection threshold predicted in Ref\,\cite{mangles12}. The second period lengthens due to the relativistic $\gamma$-factor of the electrons associated with the large amplitude plasma wave. As $n_{\rm e}$ is increased beyond $1.7\times10^{19}\,$cm$^{-3}$, $\lambda_p$ decreases but the rate of expansion increases so that we observed an approximately constant wavelength at this fixed position.

In addition, three dimensional particle in cell (3D-PIC) simulations were performed with the code EPOCH\,\cite{epoch}. A laser pulse with $\tau_{\rm L}=36\,\mathrm{fs}$ duration and $\lambda_L=810\,\mathrm{nm}$ was focused to a spot size of $18.8\,\mathrm{\mu m}$ (FWHM), $300\,\mu \rm{m}$ into a plasma density profile similar to the experiment (peak density $n_{\rm e} = 1.7\times10^{19}\,$cm$^{-3}$). In order to take into account imperfections in the experimental focal spot we set the maximum intensity of the laser pulse to $I_0=2.5\times10^{18}\,\mathrm{Wcm^{-2}}$, leading to similar energies within the focal spot FWHM in experiment and simulation\,\cite{genoud13}. The computational domain was a `sliding window' of size $150\times70\times70\,\mathrm{\mu m}^3$ moving at $c$. We used $2700\times525\times525$ cells with two electrons per cell and a stationary ion background. A $6^{th}$ order finite-difference-time-domain scheme was employed, together with $5^{th}$ order particle weighting. 
\begin{figure}[h]
\begin{center}
\includegraphics[width=246pt]{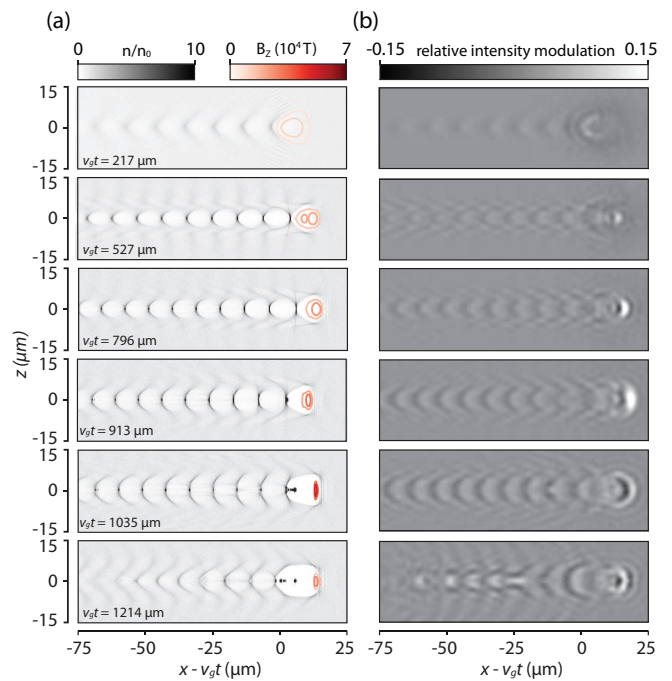}
\caption{\label{fig3}(a) Electron density maps (gray) from the simulations at various positions and contours of the envelope of pump magnetic field $B_{\rm{z}}$ (red or solid lines) corresponding to $50\%$ and $75\%$ of the peak. (b) Shadowgrams simulated from the images in Fig.\,\ref{fig3}(a).}
    \end{center}
  \end{figure}
Probe propagation has also been fully simulated in 3D with EPOCH. At different time steps during the driver pulse propagation the moving window was stopped and the probe was injected from the side of the box, propagating in the negative $y$ direction, perpendicularly to the pump. The probe had a central wavelength $\lambda_{\rm{probe}}=750\,\rm{nm}$, a Fourier limited duration of $4.4\,\mathrm{fs}$ and a negative linear chirp increasing its duration to $12\,\mathrm{fs}$ for a best match to the experimental images. The probe propagated past the wakefield structure, until $y\approx-15\,\mu$m. Subsequently, propagation in vacuum was assumed and modeled in Fourier space including the imaging system aperture, sensor sensitivity and image plane position. To adjust for the latter, we propagated the probe pulse backwards and recorded the time-averaged Poynting flux through the object plane, which  was  at $y=0$. 
The comparison of density maps and PIC generated shadowgrams in Fig.\,\ref{fig3} verifies that shadowgrams capture local variations in plasma density and allows a direct interpretation of the experimental shadowgrams. We note that the injected electron bunch appears neither in the simulated nor the experimental shadowgrams. This is caused by the reduction of local plasma frequency due to the $\gamma$-factor to the index of refraction. These simulated probe images confirm that few-femtosecond shadowgraphy provides quantitative information about the plasma wave including the plasma wave length, curvature and number of trailing periods.
\begin{figure}[h]
\begin{center}
\includegraphics[width=246pt]{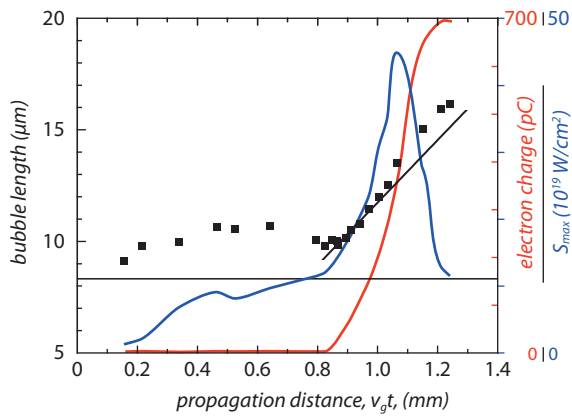}
\caption{\label{fig4}Bubble length derived from density maps (black squares) and injected charge with energy larger than 20~MeV (red or light gray) from the PIC simulation and maximum instantaneous value $S_{\rm{max}}$ of the Poynting vector magnitude (blue or dark grey). Black horizontal line is the expected $\lambda_p$ for $n_{\rm e} = 1.7\times10^{19}\,$cm$^{-3}$.}
    \end{center}
  \end{figure}

The length of the first plasma period, i.e. the bubble, as taken from the PIC simulations at various positions during the evolution is shown in Fig.\,\ref{fig4} together with the maximum amplitude of the pump pulse's Poynting vector and the total injected charge with energy above $20\,\mathrm{MeV}$. The intensity increase due to pump  compression and self-focusing [see also Fig.\,\ref{fig3}(a)]  is slow until approximately $v_{\rm g}t =800\,\mathrm{\mu m}$. Until that point the bubble length grows from $1.2\,\lambda_p$ to $1.5\,\lambda_p$. After $v_{\rm g}t =800\,\mathrm{\mu m}$ a phase of rapid intensity amplification begins, accompanied by bubble expansion and injection. At this stage, there is no substantial charge injected and so the bubble expansion is not due to beam-loading but due to intensity amplification of the pump  and the associated increase of the $\gamma$-factor of the plasma electrons; $\lambda_{\rm p}^\ast\approx\lambda_{\rm p}(1+a_0^2/2)^{1/4}$. This intensity amplification manifests itself in an increased visibility of the front of the bubble in the simulation, Fig.\,\ref{fig3}(b) and experimental (Fig.\,\ref{fig1}(e-f)) shadowgrams. In the PIC simulation significant charge is only injected into the wake (around $v_{\rm g}t =930\,\mathrm{\mu m}$) after the length of the bubble has started to increase as also observed previously in simulations\,\cite{bulanov98,mangles04,kalmykov09}. After $v_{\rm g}t =1000\,\mathrm{\mu m}$, the intensity decreases, while the length of the bubble keeps increasing due to beam-loading. At approximately $v_{\rm g}t =1300\,\mathrm{\mu m}$, the simulation shows a merging of the first two periods of the wake. The simulation supports the experimental observation that self-injection in the LWFA is caused by the expansion of the bubble. It predicts an expansion velocity of the bubble length of $(4.2\pm 0.4)\times 10^6\, \mathrm{ms}^{-1}$ (cf. Fig.\,\ref{fig4}), which is faster than the experimentally measured value. This could be due to imperfections in the experimental pulse profile leading to less efficient pulse self-compression and self-focusing than in the simulation. Our ability to determine the rate of expansion of the bubble from the experimental shadowgrams could allow the benchmarking and further development of dynamic bubble models in the future.

Applying our approach to visualize the full non-linear evolution of the plasma wave allows the acceleration process to be studied with unprecedented precision. As well as providing greater understanding of acceleration in the bubble-regime, our technique can easily be adopted to more complex acceleration geometries, e.g.\,staging\,\cite{gonsalves11}, or for beam-driven acceleration\,\cite{blumenfeld07,caldwell09,litos14}. Furthermore, increasing the probe pulse's wavelength while keeping a few-cycle duration will increase the sensitivity of our technique to probe low-density plasmas at the same relative resolution. Such low plasma densities are essential for high-energy plasma-acceleration scenarios. As LWFAs are widely expected to become useful sources of ultra short radiation\,\cite{albert14}, the increased level of understanding of plasma wave evolution and injection that can be gleaned from few-femtosecond shadowgraphy and the improvements in beam properties resulting from this technique are therefore likely to have a large impact on biomedical imaging and ultrafast condensed-matter study.

We thank B. Beleites, W. Ziegler, and F. Ronneberger for running the JETI-laser system. E.S. and S.S. thank V.~T.~Tikhonchuk for helpful discussions. This study was supported by DFG (grants TR18 A12, B9 and KA 2869/2-1), BMBF (contracts 05K10SJ2 and 03ZIK052), European Regional Development Fund (EFRE), STFC (ST/J002062/1) and EPSRC (EP/H00601X/1). The collaboration was funded by LASERLAB-EUROPE (grant agreement n$^\circ$ 284464, EC's Seventh Framework Programme). EPOCH was developed under UK EPSRC grants EP/G054940/1, EP/G055165/1 and EP/G056803/1. For the computations, the Max Planck supercomputer at the Garching Computing Center RZG was used.


\begin{thebibliography}{32} 
   
\bibitem{pukhov02} A. Pukhov and J. Meyer-ter-Vehn, Appl. Phys. B \textbf{74}, 355 (2002).
  
\bibitem{leemans06} W. P. Leemans \textit{et al.}, Nature Phys. \textbf{2}, 696 (2006).
  
\bibitem{wang13} X. Wang \textit{et al.}, Nature Commun. \textbf{4}, 1988 (2013).
  
\bibitem{lundh11} O. Lundh \textit{et al.}, Nature Phys. \textbf{7}, 219 (2011).
  
\bibitem{buck11} A. Buck \textit{et al.}, Nature Phys. \textbf{7}, 543 (2011).

\bibitem{schnell12} M. Schnell \textit{et al.}, Phys. Rev. Lett. \textbf{108}, 075001 (2012).
  
\bibitem{plateau12} G. R. Plateau \textit{et al.}, Phys. Rev.  Lett. \textbf{109}, 064802 (2012).

\bibitem{hooker13} S. M. Hooker, Nature Photon. \textbf{7}, 775 (2013).

\bibitem{mangles04} S. P. D. Mangles \textit{et al.}, Nature \textbf{431}, 535 (2004).

\bibitem{geddes04} C. G. R. Geddes \textit{et al.}, Nature \textbf{431}, 538 (2004).

\bibitem{faure04} J. Faure \textit{et al.}, Nature \textbf{431}, 541 (2004).

\bibitem{gonsalves11} A. J. Gonsalves \textit{et al.}, Nature Phys. \textbf{7}, 862 (2011).
  
\bibitem{rousse04} A. Rousse \textit{et al.}, Phys. Rev. Lett. \textbf{93}, 135005 (2004).
  
\bibitem{kneip10} S. Kneip \textit{et al.}, Nature Phys. \textbf{6}, 980 (2010).

\bibitem{powers14} S. M. Powers \textit{et al.}, Nature Photon. \textbf{8}, 28 (2014).

\bibitem{thomasPRL07} A. G. R. Thomas \textit{et al.}, Phys. Rev. Lett. \textbf{98}, 095004 (2007).

\bibitem{faurePRL05} J. Faure \textit{et al.}, Phys. Rev. Lett. \textbf{95}, 205003 (2005).

\bibitem{tzoufras09} M. Tzoufras \textit{et al.}, Phys. Plasmas \textbf{16}, 056705 (2009).

\bibitem{kalmykov09} S. Kalmykov \textit{et al.}, Phys. Rev. Lett. \textbf{103}, 135004 (2009).

\bibitem{kostyukov09} I. Kostyukov \textit{et al.}, Phys. Rev. Lett. \textbf{103}, 175003 (2009).

\bibitem{lehe13} R. Lehe \textit{et al.}, Phys. Rev. Lett. \textbf{111}, 085005 (2013).
      
\bibitem{marques96} J. R. Marques \textit{et al.}, Phys. Rev. Lett. \textbf{76}, 3566 (1996).
  
\bibitem{siders96} C. W. Siders \textit{et al.}, Phys. Rev. Lett. \textbf{76}, 3570 (1996).
  
\bibitem{matlis06} N. H. Matlis, \textit{et al.}, Nature Phys. \textbf{2}, 749 (2006).

\bibitem{Li14} Z. Li \textit{et al.}, Phys. Rev. Lett. \textbf{113}, 085001 (2014).
   
\bibitem{schwab13} M. B. Schwab \textit{et al.}, Appl. Phys. Lett. \textbf{103}, 191118 (2013).
  
\bibitem{thomas07} A. G. R. Thomas \textit{et al.}, Phys. Rev. Lett. \textbf{98}, 054802 (2007).

\bibitem{bulanov97} S. V. Bulanov \textit{et al.}, Phys. Rev. Lett. \textbf{78}, 4205 (1997).    

\bibitem{mangles12} S. P. D. Mangles \textit{et al.}, Phys. Rev. ST Accel. Beams \textbf{15}, 011302 (2012).

\bibitem{epoch} http://ccpforge.cse.rl.ac.uk/gf/project/epoch  

\bibitem{genoud13} G. Genoud, \textit{et al.}, Phys. Plasmas, \textbf{20} 064501 (2013).
   
\bibitem{bulanov98} S. Bulanov \textit{et al.}, Phys. Rev. E \textbf{58}, R5257 (1998).
   
\bibitem{blumenfeld07} I. Blumenfeld \textit{et al.}, Nature \textbf{445}, 741 (2007).
  
\bibitem{caldwell09} A. Caldwell \textit{et al.}, Nature Phys. \textbf{5}, 363 (2009).

\bibitem{litos14} M. Litos \textit{et al.}, Nature \textbf{515}, 92 (2014).

\bibitem{albert14} F. Albert, \textit{et al.}, Plasma Phys. Control. Fusion \textbf{56}, 084015 (2014).

\end{thebibliography}
\end{document}